\documentclass[runningheads]{llncs}
\usepackage{graphicx}
\usepackage{hyperref,xcolor}
\hypersetup{
    colorlinks = true,
    linkcolor = {blue},
    citecolor = {blue},
}

\usepackage{mathtools}
\usepackage{listings}
\usepackage{proof}
\usepackage{amsfonts}
\usepackage{amssymb}
\usepackage{stmaryrd}

\begin{document}
\title{Formalism-Driven Development \\ of Decentralized Systems
\thanks{To appear in ICECCS 2022.}
}
%
%
\author{Yepeng Ding \and
Hiroyuki Sato}
\authorrunning{Y. Ding and H. Sato}
%
\institute{The University of Tokyo, Tokyo, Japan\\
\email{\{youhoutei,schuko\}@satolab.itc.u-tokyo.ac.jp}}
\maketitle              
\begin{abstract}
Decentralized systems have been widely developed and applied to address security and privacy issues in centralized systems, especially since the advancement of distributed ledger technology. However, it is challenging to ensure their correct functioning with respect to their designs and minimize the technical risk before the delivery. Although formal methods have made significant progress over the past decades, a feasible solution based on formal methods from a development process perspective has not been well developed. In this paper, we formulate an iterative and incremental development process, named formalism-driven development (FDD), for developing provably correct decentralized systems under the guidance of formal methods. We also present a framework named Seniz, to practicalize FDD with a new modeling language and scaffolds. Furthermore, we conduct case studies to demonstrate the effectiveness of FDD in practice with the support of Seniz.

\end{abstract}
\section{Introduction}
Decentralization has become a ubiquitous concept in system design and implementation over the past decades such as decentralized routing protocols (e.g., RIP, OSPF) and peer-to-peer networks \cite{schollmeier_definition_2001}. To date, decentralization is still undergoing intense study in information and communications technology (ICT) and has evolved into a new stage with the advent of blockchain. The blockchain was first introduced as the underlying technology of a decentralized payment system named Bitcoin \cite{nakamoto_bitcoin_2019}. Later, it was extended by the smart contract \cite{buterin_next-generation_2014} and generalized to a concept named distributed ledger technology (DLT). Based on the DLT, numerous decentralized systems have been developed to address security and privacy issues in a wide range of fields such as the Internet of Things (IoT), data persistence, and security infrastructure.

The popularity of the DLT can be credited to its attractive characteristics such as immutability, fault tolerance, non-repudiation, transparency, traceability, and auditability. Without central authorities, data, one of the most valuable assets nowadays, is protected from vulnerabilities and threats lying in centralized systems by decentralizing central entities and returning the ownership to data owners. However, it is neither for free nor a silver bullet. In fact, it is incredibly challenging to develop a trustworthy decentralized system that preserves these characteristics \cite{fu_evmfuzzer_2019} even for experienced architects and developers. For instance, Geth, the most widely used implementation of Ethereum virtual machine (EVM), still has vulnerabilities that lead to consensus errors. The most recent one \footnote{https://github.com/ethereum/go-ethereum/security/advisories/GHSA-9856-9gg9-qcmq} potentially causing a node to reject the canonical chain has led to a minority chain split after the London hard fork. These vulnerabilities are hard to locate but can bring critical security issues \cite{yang_finding_2021} by threatening core mechanisms such as the consensus mechanism.

When it comes to system correctness, formal methods have proved to be effective in the specification, verification, and testing, such as model checking \cite{holzmann_model_1997} and theorem proving \cite{paulson_isabelle_1994}. Formal specification rigorously describes system behaviors and constrains implementations, while formal verification proves correctness and system properties with respect to specifications. Furthermore, the application of formal methods for developing provably correct decentralized systems ranging from blockchain platforms \cite{hildenbrandt_kevm_2018} to smart contracts \cite{bhargavan_formal_2016} has drawn widespread attention both in academia and industry since the DAO attack.

However, these works either verify systems at the implementation (code) level or verify models extracted from implementations. In the first case, implementation verification focuses on the details of the execution, such as runtime bugs (e.g., null pointer, division by 0, buffer overflow), functional correctness bugs (e.g., undefined behaviors, unexpected algorithm output), and concurrency bugs (e.g., deadlock, race condition). It is crucial to verify implementations and even worthwhile analyzing bytecodes. Nevertheless, the implementation verification is a unilateral strategy and hard to unravel the design flaw. In the second case, it is possible to locate design flaws in a system by verifying specified properties of models. However, extracting a proper model from the complicated implementation is nontrivial. Extracted models might be too simple to have useful properties or too complicated to be verified \cite{valmari_state_1996}. Besides, it is hard to judge whether a model extracted from an implementation coincides with its design. If there is a departure from the design, properties associated with that model can be untrustworthy and meaningless, which we call a conformity issue. Furthermore, it still lacks a standardized development process \cite{destefanis_smart_2018} for developing decentralized systems with formal methods in a practical and usable way.

In this paper, we propose a novel iterative and incremental development process called Formalism-Driven Development (FDD) for developing provably correct decentralized systems. Our motivation is to address the issues above that hinder the development of trustworthy decentralized systems. We tackle these issues from the perspective of a development process that we regard as the root cause. Our core methodology effectively integrates formal methods throughout the whole development lifespan, which suits our objective: to produce rigorous designs, mathematically verifiable models, and provably correct implementations. We summarize our main contributions as follows.

\begin{enumerate}
    \item We formulate FDD by introducing a new formalism to model decentralized systems, facilitate formal verification, and generate robust skeleton codes.
    \item Based on the theoretical foundations of FDD, we show a framework named Seniz that practicalizes FDD.
    \item We conduct case studies to demonstrate the effectiveness of adopting FDD with Seniz in real-world projects.
\end{enumerate}

\section{Preliminaries}
In this section, we introduce core concepts and theories associated with FDD.

\subsection{Basic Structures}
In this paper, we define a variant of the labeled transition system as below.

\begin{definition}[Labeled Transition System]
\label{def:lts}
A labeled transition system $\mathfrak{T}$ over set \textit{Var} of typed state variables is a tuple
\begin{equation*}
    \mathfrak{T} \triangleq \langle S, A, {\to}, I, P, \mathcal{L} \rangle
\end{equation*}

where
\begin{itemize}
    \item $S = \llbracket \textit{Var} \rrbracket$ is a set of states,
    \item $A$ is a set of actions,
    \item ${\to} \subseteq S \times A \times S$ is a transition relation,
    \item $I \subseteq S$ is a set of initial states,
    \item $P$ is a set of atomic propositions, and
    \item $\mathcal{L}: S \mapsto \wp(P)$ is a labeling function.
\end{itemize}
The state space $S$ is determined by $\llbracket \textit{Var} \rrbracket$, the set of evaluations of state variables \textit{Var}. State $s \in S$ is called a terminal state if it does not have any outgoing transitions, i.e., $\nexists a \in A: s \xrightarrow{a} s'$. The notation $s \xrightarrow{a} s'$ is used as shorthand for $(s, a, s') \in {\to}$. In this paper, we assume that $S$, $A$, and $P$ are finite sets.
\end{definition}

In the remainder of this paper, we abbreviate \textit{labeled transition system} to \textit{transition system}.

Conditional branching is commonly used in modeling systems. By using conditional branching, it is possible to put constraints on actions. An action can only be triggered while the current evaluation of variables satisfies some conditions. We denote a set of Boolean conditions (propositional formulae) over \textit{Var} as $\| \textit{Var} \|$. In the interest of modeling conditional branching, we introduce conditional transitions.

\begin{definition}[Conditional Transition]
\label{def:cond_trans}
A transition system $\mathfrak{T}$ with conditional transitions over set $\textit{Var}$ of typed state variables is a tuple
\begin{equation*}
    \mathfrak{T} \triangleq \langle S, A, {\hookrightarrow}, I, g_0, P, \mathcal{L} \rangle
\end{equation*}
according to Definition~\ref{def:lts} with differences that
\begin{itemize}
    \item ${\hookrightarrow} \subseteq S \times \| \textit{Var} \| \times A \times S$ is the conditional transition relation, and
    \item $g_0 \in \| \textit{Var} \|$ is the initial guard (condition).
\end{itemize}
For convenience, we use the notation $s \xhookrightarrow{g \downarrow a} s'$ as shorthand for $(s, g, a, s') \in {\hookrightarrow}$. If the guard is a tautology, we can omit it, i.e., $s \xhookrightarrow{a} s'$.

The behavior in state $s \in S$ depends on the current state variable evaluation $\mathcal{V} \in \llbracket \textit{Var} \rrbracket$. The value of state variable $x \in \textit{Var}$ is accessible through $\mathcal{V}(x)$. For transition $s \xhookrightarrow{g \downarrow a} s'$, the execution of action $a$ is only triggered when evaluation $\mathcal{V}$ satisfies guard $g$, i.e., $\mathcal{V} \models g$. The new evaluation can be represented by changed state variables, e.g., $\mathcal{V}' = \mathcal{V}[x:v]$, meaning that state variable $x$ has value $v$ in $\mathcal{V}'$ and all other state variables are unaffected.
\[  \mathcal{V}[x:v](x')= \begin{cases} 
      \mathcal{V}(x') & x \neq x' \\
      v & x = x' .
   \end{cases}
\]
\end{definition}

Given a transition system with conditional transitions, it is natural that it can be transformed into an equivalent transition system without conditional transitions.

\begin{definition}[Semantics of Conditional Transitions]
Let $\mathfrak{T} = \langle S, A, {\hookrightarrow}, I, g_0, P, \mathcal{L} \rangle$ be a transition system with conditional transitions over set \textit{Var} of typed state variables. The corresponding transition system $\mathfrak{T}'$ without conditional transitions is the tuple $\langle S, A, {\to}, I, P, \mathcal{L} \rangle$ where
\begin{itemize}
    \item ${\to}$ is defined by the following rule:
    \begin{equation*}
        \infer{s \xrightarrow{a} s'}{s \xhookrightarrow{g \downarrow a} s' \land \mathcal{V} \models g},
    \end{equation*}
    \item $I = \{ \langle s, \mathcal{V} \rangle ~|~ s \in I, \mathcal{V} \models g \}$, and
    \item $S, A, P, \mathcal{L}$ remains the same.
\end{itemize}
\end{definition}

\begin{remark}[State Tautology]
If $S=\llbracket \textit{Var} \rrbracket$, the current state $s$ and current state variable evaluation $\mathcal{V}$ are interchangeable. The tuple $\langle s, \mathcal{V} \rangle$ can be reduced to either $s$ or $\mathcal{V}$.
\end{remark}

\begin{remark}[State Rewriting]
\label{rem:state_rewriting}
If $s = \langle \mathcal{V}_1, \mathcal{V}_2, \dots, \mathcal{V}_n \rangle$ where $s \in S$ and $\bigcap\limits_{i=1}^{n}\textit{Dom}(\mathcal{V}_i) = \emptyset$, $s$ can be rewritten as a merged variable evaluation $\mathcal{V} = \bigoplus\limits_{i=1}^{n}\mathcal{V}_i$. Here, $\bigoplus$ notation is used to indicate repeated $\oplus$.
\end{remark}

\subsection{Parallelism}
\label{sec:parallelism}
Parallel systems can also be modeled by transition systems. In this paper, we introduce two common types of parallelism \cite{nielsen_models_1991}: asynchronous concurrency (pure interleaving) and synchronous concurrency (variable sharing).

Informally, asynchronous concurrency models a parallel system composed of a set of independent subsystems, i.e., the intersection of variables of subsystems is empty. And synchronous concurrency models a parallel system whose subsystems have variables in common. In this case, contentions on shared variables need to be solved. We define it as follows.

\subsection{Communication}
To model distributed systems, a communication model is indispensable. In this paper, we model the communication by channels. A channel is a buffer based on a queue where messages are stored and held to be processed later.

Given channel $c$, we define a set of functions to access the properties of $c$. $c$ has a finite capacity $\textit{Cap(c)} \in \mathbb{N}$ and a domain $\textit{Dom}(c)$. The current number of messages in $c$ is fetched by \textit{Len(c)}. Besides, we can manipulate contents of $c$ by a set of operations. $\textit{Enq}(c, m)$ puts message $m$ at the rear of the buffer whereas $\textit{Deq}(c)$ pops an element from the front of the buffer.

We introduce two actions for sending and receiving messages based on the operations of $c$.
\begin{itemize}
    \item $c!m$: send the message $m$ along channel $c$, i.e., $\textit{Enq}(c, m)$,
    \item $c?x$: receive a message via channel $c$ and variable $x$ has value of the message, i.e., $x: \textit{Deq}(c)$.
\end{itemize}

With two message-passing actions, we define the set of communication actions \textit{Com} as:
$\textit{Com} = \big\{ c!m, c?x ~|~ c \in \textit{Chan}, m \in \textit{Dom}(c), x \in \textit{Var} \text{ with } \textit{Dom}(x) \supseteq \textit{Dom}(c) \big\}$, where \textit{Chan} is a set of channels with typical element $c$ and \textit{Var} is a set of variables as in Definition~\ref{def:lts}.

\begin{definition}[Channel System]
\label{def:channel_sys}
A transition system with conditional transitions $\mathfrak{T}$ over $(\textit{Var}, \textit{Chan})$ is a tuple
\begin{center}
    $\mathfrak{T} \triangleq \langle S, A, {\hookrightarrow}, I, g_0, P, \mathcal{L} \rangle$
\end{center}
according to Definition~\ref{def:lts} and Definition~\ref{def:cond_trans} with the only difference that ${\hookrightarrow} \subseteq S \times \| \textit{Var} \| \times (A \cup \textit{Com}) \times S$.

A channel system $\mathfrak{C}$ over $(\textit{Var}, \textit{Chan}), \textit{Var}=\bigcup\limits_{i=1}^{n}\textit{Var}_i$ with $\bigcap\limits_{i=1}^{n}\textit{Var}_i = \emptyset$, consisting of transition systems $\mathfrak{T}_i$ over $(\textit{Var}_i, \textit{Chan}), i \in [1, n]$ is defined as
\begin{center}
    $\mathfrak{C} \triangleq [\mathfrak{T}_1 ~|~ \dots ~|~ \mathfrak{T}_n]$.
\end{center}
\end{definition}

\begin{remark}[State Structure of a Channel System]
\label{rem:state_chan_sys}
Let $\mathfrak{C}=[\mathfrak{T}_1 ~|~ \dots ~|~ \mathfrak{T}_n]$ be a channel system over $(\textit{Var}, \textit{Chan})$. The global states $S$ of $\mathfrak{C}$ are tuples of the form $\langle s_1, \dots, s_n, \mathcal{V}, \mathcal{C} \rangle$, where
\begin{itemize}
    \item $s_i \in S_i$ is the current state (variable evaluation) of subsystem $\mathfrak{T}_i$, i.e., $s_i = \mathcal{V}_i \in \llbracket \textit{Var}_i \rrbracket$,
    \item $\mathcal{V} = \bigoplus\limits_{i=1}^{n} \mathcal{V}_i \in \llbracket \textit{Var} \rrbracket$ is the current variable evaluation (state) of $\mathfrak{C}$, i.e., $\mathcal{V} = s \in S$, and
    \item $\mathcal{C} \in \llbracket \textit{Chan} \rrbracket$ is the current channel evaluation.
\end{itemize}

$\mathcal{C}$ is a mapping from channel $c \in \textit{Chan}$ onto a sequence $\mathcal{C}(c) \in \textit{Dom}(c)^*$ such that $\textit{Len}(\mathcal{C}(c)) \leqslant \textit{Cap}(c)$, e.g. $\mathcal{C}(c) = [v_1v_2 \dots v_k]$ with $\textit{Cap}(c) \geqslant k$. If $\textit{Cap}(c) = 0$, $c$ is a synchronous channel. Otherwise, $c$ is an asynchronous channel.
\end{remark}

Notably, a channel system can have a nested structure, i.e., a subsystem can also be a channel system or parallel system. If a channel system only contains one transition system with conditional transitions, it is merely a transition system with conditional transitions and channel extension. Therefore, a channel system is capable of describing complex structures and behaviors of a distributed system.

Furthermore, given a channel system, there exists an equivalent transition system \cite{baier_principles_2008}. The interpretation from a channel system to a transition system can be automated according to the transition system semantics of a channel system, which permits us to model a system on top of a channel system without considering the details of its underlying transition system. It is also flexible to use different models or their combinations according to concrete contexts.

\subsection{Properties}
One important reason to model a system is to facilitate specifying and studying its properties in a rigorous way. In our current work, we use temporal logic, a formalism par excellence for mathematically expressing properties about system behaviors. More concretely, we use propositional temporal logic, including linear-time and branching-time logic.

\section{Formalism-Driven Development}
\label{sec:fdd}
Formalism-driven development (FDD) is an iterative and incremental development process promoting formal methods throughout the lifespan. It is devised to take advantage of formal methods to eliminate design ambiguity, prove model properties, verify and test implementation correctness, and ensure conformity among design, model, and implementation. The core idea is to elaborate transition system theory to bond all concepts together.

In fact, the philosophy of iterative and incremental development process is widely practiced in agile development \cite{shore_art_2007}. Nevertheless, both iteration and increment are not formally defined in agile processes. Generally, iteration means enhancing systems progressively, while increment means delivering the system by pieces. However, it is hard to give a well-defined explanation about what an iteration or increment produces as well as relations between two iterations and relations between an iteration and an increment. In FDD, iteration and increment are defined based on a formalism with theory support, including modeling, refinement, and verification \cite{baier_principles_2008}. With these well-defined theories, iterations and increments can be rigorously managed and used to produce verifiable deliveries.

\begin{figure}
\centering
\includegraphics[width=9cm]{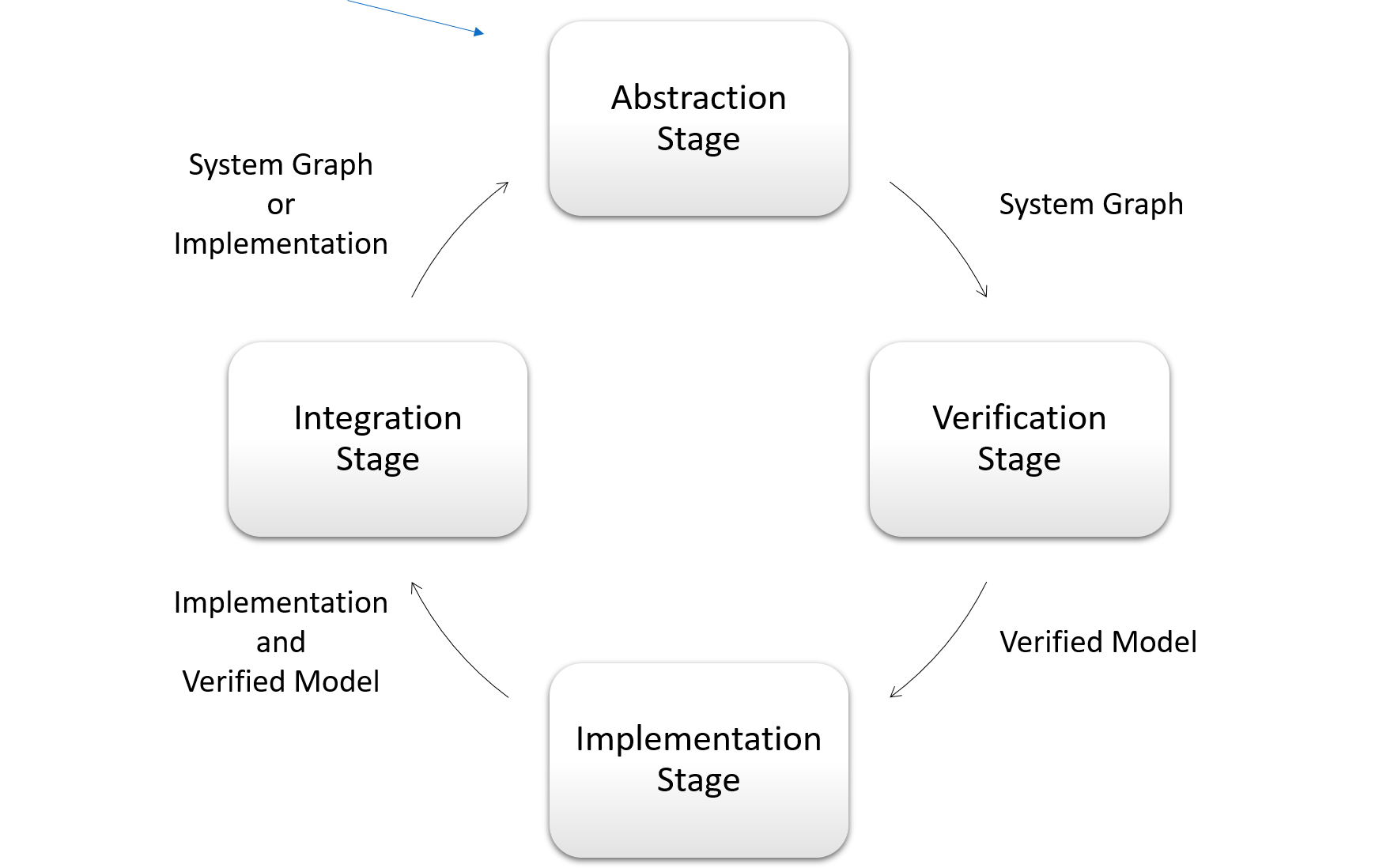}
\caption[Stage Transition Graph]{Stage transition graph of an iteration.}
\label{fig:stages}
\end{figure}

In FDD, an iteration formulates a design model, proves model properties, implements the model, verifies the model implementation, and integrates or delivers the milestone. An increment organizes subsystems together as a higher-level system. Concretely, an iteration contains four stages: abstraction, verification, implementation, and integration, which is shown in Figure~\ref{fig:stages}. \textit{Abstraction Stage} produces system graphs as design models. In \textit{Verification Stage}, system graphs are verified by formal verification. \textit{Implementation Stage} only accepts verified models to generate skeleton programs and implement concrete functionalities. In \textit{Integration Stage}, system graphs are integrated into higher-level systems or delivered. Naturally, an increment comes from \textit{Integration Stage} and can also launch a set of new iterations.

\subsection{Abstraction Stage}
In \textit{Abstraction Stage}, the goal is to produce a rigorous design model (system graph) for a system. If it is the first iteration of a new system, a model is built from the ground up, which is called \textit{Origin Stage}. Otherwise, we call it \textit{Refinement Stage} where a model from the last iteration is refined.

\subsubsection{Origin Stage}
\textit{Origin Stage} creates a system graph as a design model. A system graph is built on top of a channel system defined in Definition~\ref{def:channel_sys}. Although they have equivalent expressiveness, a system graph cuts down the details that describe individual states by using a naming function to describe a group of states. Besides, individual states are inferred from concrete contexts. This keeps a system graph succinct to model complex systems such as decentralized systems.

\begin{definition}[System Graph]
\label{def:sys_graph}
A system graph $\mathfrak{S}$ over $(\textit{Var}, \textit{Chan})$ is a tuple
\begin{center}
$\mathfrak{S} \triangleq \langle D, \mathcal{N}, A, {\hookrightarrow}, i, g_0, F, P, \mathcal{L} \rangle$
\end{center}
where
\begin{itemize}
    \item $D=N \times \llbracket \widehat{\textit{Var}} \rrbracket, \widehat{\textit{Var}} \subseteq \textit{Var}$ is a set of state declarators with names in $N$,
    \item $\mathcal{N}: D \mapsto \wp(\llbracket \textit{Var} \rrbracket)$ is a naming function,
    \item $A \supseteq \textit{Com}$ is a set of actions,
    \item ${\hookrightarrow} \subseteq D \times \| \textit{Var} \| \times A \times D$ is the conditional transition relation,
    \item $i \in D$ is the initial state declarator,
    \item $g_0 \in \| \textit{Var} \|$ is the initial guard,
    \item $F \subseteq D$ is a set of terminal state declarators,
    \item $P \supseteq \| \textit{Var} \|$ is a set of propositions, and
    \item $\mathcal{L}: \llbracket \textit{Var} \rrbracket \mapsto \wp(P)$ is a labeling function.
\end{itemize}
Notably, a system graph uses state declarators to describe state sets and infer individual states instead of identifying each state explicitly. A state declarator $d \in D$ introduces a kind of state with a given name into a system by identifying \textbf{interesting} state variables that are essential to show features of this kind of state. The name of a state declarator is unique, i.e.,
\[
\forall \langle n, \widehat{\mathcal{V}} \rangle \in D (\nexists \langle n', \widehat{\mathcal{V}}' \rangle \in D: \widehat{\mathcal{V}} = \widehat{\mathcal{V}}' \land n \neq n').
\]
The naming function $\mathcal{N}$ relates a set $\mathcal{N}(d) \in \wp(\llbracket \textit{Var} \rrbracket)$ of variable evaluations, i.e., states, to any state declarator $d=\langle n, \widehat{\mathcal{V}} \rangle$ such that
\begin{itemize}
    \item $\forall \mathcal{V} \in \llbracket \textit{Var} \rrbracket (\exists! \widehat{\mathcal{V}} \in \llbracket \widehat{\textit{Var}} \rrbracket: \mathcal{V} \in \mathcal{N}(d))$, and
    \item $\forall x \in \textit{Dom}(\widehat{\mathcal{V}}): \widehat{\mathcal{V}}(x) = \mathcal{V}(x), \mathcal{V} \in \llbracket \textit{Var} \rrbracket, \textit{Dom}(\widehat{\mathcal{V}}) \subseteq \textit{Dom}(\mathcal{V})$.
\end{itemize}

The conditional transition relation is on top of state declarators. Only one initial state declarator exists in a system graph. A system graph is nonterminal if $F = \emptyset$. Propositions are well-formed propositional formulae in propositional logic and constructed from atomic propositions by logical connectives. A set of propositions are related to any variable evaluation, i.e., state, by the labeling function $\mathcal{L}$.
\end{definition}

By using the state declarator, it enables describing a system in a succinct form. We are only concerned about the most critical features of states identified by interesting state variables. The evaluation of other state variables is inferred from the preceding state.

\begin{example}[Transaction Client]
\label{eg:tx_client}
We use a simplified transaction client in our developed demonstration as an example to illustrate core concepts in this paper. The complete demonstration fully developed by FDD is a prototype of Ethereum including the client-side and server-side systems.

A visualized system graph $\mathfrak{S}_{tx}$ of the simplified transaction client is shown in Figure~\ref{fig:tx_client}.

\begin{figure}[htbp!]
\centering    
\includegraphics[width=9cm]{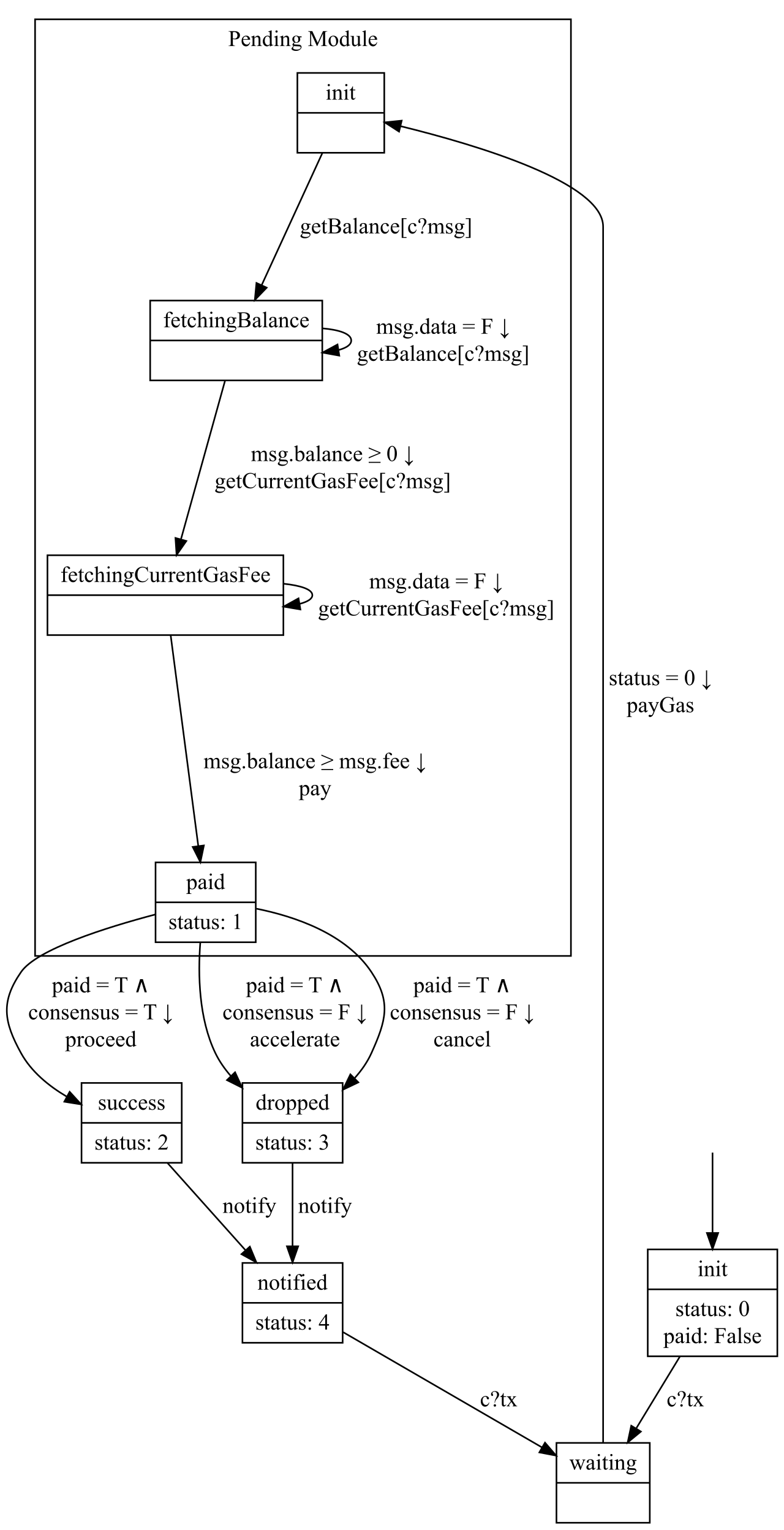}
\caption[Visualized System Graph of Transaction Client]{Visualized system graph of the transaction client in Example~\ref{eg:tx_client}.}
\label{fig:tx_client}
\end{figure}

Each box is a state declarator consisting of two parts. The big box is also an increment (illustrated in Section~\ref{sec:integration_stage}). The above part is a name, while the below part is a set of state variable evaluations. State variable \textit{status} has type Integer and \textit{paid} has type Boolean. A one-way arrow pointing from a state declarator to another is a transition relation with a guard and an action. The initial state declarator is pointed by an arrow without the starting node. We omit the representation of the guard if it is a tautology, e.g., the initial guard.

\end{example}

\begin{definition}[State Inference]
\label{def:state_inference}
Let $\mathfrak{S}$ be a system graph over $(\textit{Var}, \textit{Chan})$. The state space $S$ of $\mathfrak{S}$ is determined by $\llbracket \widehat{\textit{Var}} \rrbracket \times \llbracket \textit{Var} \setminus \widehat{\textit{Var}} \rrbracket$.

Let $\mathcal{V} \in \llbracket \textit{Var} \rrbracket$ be the current variable evaluation. According to Remark~\ref{rem:state_rewriting} and Definition~\ref{def:cond_trans}, the succeeding state $s' \in S$ named by a state declarator $d' = \langle n', \widehat{\mathcal{V}}' \rangle \in D$ is represented as a variable evaluation $\mathcal{V}' = \langle \widehat{\mathcal{V}}', \mathcal{V} \ominus \widehat{\mathcal{V}}' \rangle = \mathcal{V}[\widehat{\mathcal{V}}']$ such that
\[   \forall x \in \textit{Var}: \mathcal{V}[\widehat{\mathcal{V}}'](x)= \begin{cases} 
      \widehat{\mathcal{V}}'(x) & x \in \textit{Dom}(\widehat{\mathcal{V}}') \\
      \mathcal{V}(x) & x \notin \textit{Dom}(\widehat{\mathcal{V}}').
    \end{cases}
\]
The initial state $s_0$ named by the initial state declarator $i$ is represented as a variable evaluation $\mathcal{V}_0 = \mathcal{V}[i]$ such that
\[   \forall x \in \textit{Var}: \mathcal{V}[i](x)= \begin{cases} 
      i(x) & x \in \textit{Dom}(i) \\
      \epsilon & x \notin \textit{Dom}(i)
   \end{cases}
\]
where $\epsilon$ denotes the default value.
\end{definition}

\begin{example}[State Inference in Transaction Client]
In Example~\ref{eg:tx_client}, state declarator named \textit{waiting} does not identify any interesting state variables. According to Definition~\ref{def:state_inference}, the evaluation of state variables \textit{status} and \textit{paid} for state declarator \textit{waiting} remains the same with \textit{init}, i.e., \textit{status} has value $0$ and \textit{paid} has value \textit{False}. But \textit{waiting} is distinguished from \textit{init} by a hidden state variable \textit{tx}. State declarator \textit{waiting} implies state variable \textit{tx} has value of the first element in channel $c$.
\end{example}

\begin{remark}[Transition Interpretation]
\label{rem:trans_interpret}
In Definition~\ref{def:state_inference}, states of $\mathfrak{S}$ are inferred from contexts. Correspondingly, the declarator-based conditional transition relation $\hookrightarrow$ of $\mathfrak{S}$ is interpreted to a state-based conditional transition relation by the following rule:
\[
    \infer{\langle \widehat{\mathcal{V}}, \mathcal{V} \ominus \widehat{\mathcal{V}}' \rangle \xhookrightarrow{g \downarrow a} \langle \widehat{\mathcal{V}}', \mathcal{V}' \ominus \widehat{\mathcal{V}}' \rangle}{d \xhookrightarrow{g \downarrow a} d' \land \mathcal{V} \models g}.
\]
\end{remark}

\begin{remark}[Transition System Semantics of a System Graph]
\label{rem:sys_graph_semantics}
Let $S$ over $(\textit{Var}, \textit{Chan})$ be a system graph $\langle D, \mathcal{N}, A, {\hookrightarrow}, i, g_0, F, P, \mathcal{L} \rangle$. According to Definition~\ref{def:sys_graph}, Definition~\ref{def:state_inference}, and Remark~\ref{rem:trans_interpret}, $\mathfrak{S}$ can be transformed into a channel system $\mathfrak{C}$ with the requirement that
\begin{itemize}
    \item $\forall p \in P:$ $\textit{Atom}(p)$ are contained in the atomic proposition set of $\mathfrak{C}$,
    \item $\forall s \in \llbracket \textit{Var} \rrbracket: \forall p \in \mathcal{L}(s):$ $\textit{Atom}(p)$ are related to $s$ through the labeling function of $\mathfrak{C}$,
\end{itemize}
where $\textit{Atom}(p)$ denotes all atomic propositions contained in the conjunctive normal form of $p$.

By the transition system semantics of a channel system, $\mathfrak{C}$ can be interpreted over a transition system.

Notably, the termination of a system graph does not imply the termination of its underlying transition system, and vice versa. $F$ is omitted during the interpretation.
\end{remark}

By interpreting a system graph over a transition system, we can use the high-level design model, system graph, to model systems while safely using transition system theory to support prominent features of FDD in later stages and iterations.

\subsubsection{Refinement Stage}
\textit{Refinement Stage} accepts a system graph from the last iteration and produces a more detailed system graph while preserving and extending properties. According to Remark~\ref{rem:sys_graph_semantics}, a system graph can be interpreted onto a transition system. In FDD, we use both bisimulation and simulation theory \cite{milner_algebraic_1971,groote_efficient_1990} to support the refining process. The original purposes of these techniques are generally to optimize the verification process and improve verification efficiency by compacting a model while preserving its properties. However, \textit{Refinement Stage} inverses the original purpose to extend a small model into a big one while preserving its properties.

A refined system graph (refinement) $\mathfrak{S}'$ of $\mathfrak{S}$ is a more detailed design model that has either a strong relation $\sim$ or a weak relation $\preceq$ to $\mathfrak{S}$. Relation $\sim$ is an equivalence relation that identifies $\mathfrak{S}$ and its refinement with the same branching structure by bisimulation. Relation $\preceq$ is a preorder. $\mathfrak{S}’ \preceq \mathfrak{S}$ holds if the refinement $\mathfrak{S}’$ can be simulated by $\mathfrak{S}$.

\begin{example}[Refinement in Transaction Client]
\label{eg:tx_client_refined}
Accelerating service is included in our transaction client as an additional branch. Notably, the transaction client needs to cancel the original transaction firstly and resend a new transaction with more gas due to the immutability of the blockchain. Consequently, the current transaction is still dropped by the network.

Let $\mathfrak{S}_{tx}'$ be the system graph without \textit{accelerate} branch. We can prove that $\mathfrak{S}_{tx}$ and $\mathfrak{S}_{tx}'$ are bisimulation-equivalent with bisimulation techniques.
\end{example}

By using bisimulation and simulation techniques, properties of an original system graph can be well preserved in the next iteration if the refined system graph passes either bisimilarity or similarity verification. While encountering a violation, it allows flexible handling methods. If the properties of the original system graph are finalized, then the refinement needs to be modified until passing the verification. It is also a solution to delegate the violation to the \textit{Verification Stage} and resolve it by optimizing old properties.

\subsection{Verification Stage}
\label{sec:verify_stage}
\textit{Verification Stage} produces a verifiable model based on the input system graph by specifying the admissible behaviors of the system graph as properties. Besides, it verifies properties associated with the verified model by formal verification.

\subsubsection{Specification}
According to Definition~\ref{def:sys_graph}, a system graph has a set of propositions. Based on these propositions, we can specify essential system behaviors as a set of properties with temporal logic such as linear temporal logic and computation tree logic.

\begin{example}[Linear-Time Property in Transaction Client]
\label{eg:tx_client_ltp}
To verify whether the transaction client in Example~\ref{eg:tx_client} infinitely often gets notified after paying gas, we firstly define propositions $\textit{Notified} \triangleq \textit{status} = 4$ and $\textit{PaidGas} \triangleq \textit{status} = 1$ with respect to system graph $\mathfrak{S}_{tx}$.

Then we can formally specify the property as $\Box (\textit{PaidGas} \to \Diamond \textit{Notified})$. By the labeling function $\mathcal{L}_{tx}$ of $\mathfrak{S}_{tx}$, the states are automatically labeled with corresponding propositions. In this manner, the satisfiability of the property can be verified on the transition system under $\mathfrak{S}_{tx}$.

\end{example}

\subsubsection{Enforcement}
The enforcement of the verification depends on the verification mode of FDD. Either a model checker or a theorem prover can be used to prove properties formulated in some logic.

\paragraph{Checker Mode}
By structuring a system graph and interpreting it over a transition system, it is trivial to enforce model checking to verify the properties.

\paragraph{Prover Mode}
Properties of a system graph are verified by a theorem prover that mechanizes the logic used to specify these properties.

\subsection{Implementation Stage}
\label{sec:impl_stage}
In \textit{Implementation Stage}, a skeleton program is generated from the verifiable model. Based on the skeleton program, a real-world program is implemented with full functionality. Besides, a formal verification and testing process is enforced to ensure implementation correctness.

A skeleton program generated from the verifiable model has strict constraints to ensure conformity between the model and implementation at best efforts. The smallest skeleton program includes
\begin{enumerate}
    \item predefined and immutable state variables,
    \item predefined and immutable deterministic control flow, and
    \item predefined and overridable action effects.
\end{enumerate}
The term \textbf{predefined} means that the modificand is generated before the manual implementation. Something that can only be accessed but cannot be changed by implementors is \textbf{immutable}. An \textbf{action effect} of action $a$ is the actual functionality produced every time executing $a$ in the control flow. By overriding an action effect with an effect function, implementors can implement specific functionalities such as executing an algorithm, interacting with an I/O stream, etc.

\begin{example}[Action Effect]
In Example~\ref{eg:tx_client}, all actions produce corresponding action effects such as $c?\textit{tx}$, \textit{payGas}, \textit{proceed}, and \textit{notify}. Each of them is overridable. For instance, a logging function can be called in each of them to write the current timestamp and action name into a local file system.
\end{example}

Depending on the programming paradigm, generated skeleton programs are different on the code level. In this paper, we illustrate possible generation methods with respect to two mainstream programming paradigms and key points.

\subsubsection{General Skeleton}
\paragraph{Object-Oriented Programming}
The typical features of a system graph are extracted and formed as an abstract class $\mathbf{A}_{sys}$ that defines protected methods associated with the control flow and exposes an entry point to execute the system.

For a system graph $\mathfrak{S} = \langle D, \mathcal{N}, A, {\hookrightarrow}, i, g_0, P, \mathcal{L} \rangle$ over $(\textit{Var}, \textit{Chan})$, it contains all the information to create an abstract class $\mathbf{A}_{sg}$ that is capable of fully describing $\mathfrak{S}$. $\mathbf{A}_{sg}$ inherits $\mathbf{A}_{sys}$ and implements at least interface $\mathbf{I}_{act}$ that contains a set of method signatures extracted from all manually labeled actions in $A$. $\mathbf{A}_{sg}$ also overrides the control flow according to $D, {\hookrightarrow}, i, g_0$. $\mathbf{A}_{sg}$ together with all its associated classes forms the smallest class set (skeleton program) to describe $\mathfrak{S}$. The skeleton program is encapsulated into a package as a software development kit (SDK).

With such an SDK, implementors can create a concrete class $\mathbf{C}$ that inherits $\mathbf{A}_{sg}$. The implementors can override the effect methods (methods declared in $\mathbf{A}_{sg}$) to implement the functionality. Notably, implementors can neither modify state variables nor change the deterministic control flow. In this manner, the verified properties in \textit{Verification Stage} are preserved in the executable system.

\paragraph{Functional Programming}
In fact, it is straightforward to construct a system graph in functional programming. Related definitions can be easily formulated with customized data types. The impure action effects are isolated by the monad. All components are packaged into a module as a library that exposes a set of functions taking effect functions as their parameters and the entry point. The implementors can implement functionalities by passing the implementation of effect functions into the exposed functions.

\subsubsection{Termination}
A system graph is terminated if $F \neq \emptyset$. The execution naturally terminates while reaching a terminal state declarator or a terminal state of its underlying transition system defined in Definition~\ref{def:lts}. Without considering exception handling, the execution generated from a nonterminal system graph interpreted over a nonterminal transition system will never naturally terminate such as the transaction client in Example~\ref{eg:tx_client}.

\subsubsection{Parallelism}
In Section~\ref{sec:parallelism}, we present two types of parallelism: pure interleaving and variable sharing. Both present nondeterminism during the actual execution. The skeleton program handles them by multithreading techniques. Each system graph is encapsulated into a thread. In this manner, the implementation of nondeterminism in a parallel system is delegated to nondeterminism in thread scheduling.

\subsubsection{Divergence and Confluence}
A system graph may contain nondeterministic transitions after being interpreted over a transition system with conditional transitions. For a state $s$ with a set ${\hookrightarrow}_{out}^s$ of outgoing transitions, if there exist at least two transitions $s \xhookrightarrow{g \downarrow a} s_1, s \xhookrightarrow{g’ \downarrow a’} s_2 \in {\hookrightarrow}_{out}^s$ where $g \implies g’ \land a \neq a' \land s_1 \neq s_2$, then it is a nondeterministic choice, which is called a divergence.

\begin{example}[Divergence in Transaction Client]
\label{eg:tx_client_divergence}
In our transaction client, two transition relations from \textit{pending} to \textit{dropped} form a divergence because the truth values of their guards are the same.
\end{example}

To resolve a divergence, an \textbf{interactive event} is emitted to wait for a signal that determines a choice to resume the execution in that branch. An interactive event can be user input via I/O stream, in-memory or on-disk interaction with another program, communication through a network protocol, etc. The skeleton program exposes all divergences in the form of interfaces that need to be implemented as interactive events by implementors. While encountering a divergence, the execution pauses until getting a signal from the interactive event to proceed.

\begin{example}[Divergence Resolution for Transaction Client]
To resolve the divergence in Example~\ref{eg:tx_client_divergence}, we can use keyboard event as the interactive event by implementing a keyboard listener in a local environment for test. For instance, the effect of action \textit{cancel} is triggered while getting an input sequence \textit{c\textbackslash r\textbackslash n}.

In our demonstration, the transaction client is developed as a mobile application. Action effects are triggered by touching corresponding buttons in the UI.
\end{example}

Confluence is usually not an interesting problem because the state inference in Definition~\ref{def:state_inference} eliminates nondeterminism of implicit state variables during the execution. One exception is for a set of systems to be confluent in a parallel system that contains nondeterminism in implementation. If a parallel system has terminal states, then we say this parallel system is naturally confluent. Each terminal state is a confluence where nondeterminism is eliminated. For a parallel system without terminal states, it allows implementors to customize the confluence where all threads join by manually identifying the evaluation of state variables in that confluence.

\subsubsection{Channel}
According to Remark~\ref{rem:state_chan_sys}, a channel can be either synchronous or asynchronous. A synchronous channel usually serves synchronization purposes instead of data transfer within a system modeled by a system graph. Its data structure at least contains the metadata. For an asynchronous channel, it contains at least the metadata, a buffer, and a set of operations associated with the buffer.

Regarding the implementation, a channel has two types: internal channel and external channel. An \textbf{internal channel} only receives messages within the system while an \textbf{external channel} can also receive messages from the outside of the system. An internal channel is naturally embedded into the control flow, while an external channel requires interaction with processes outside of the system. An outside process can be a program that sends messages to channels (by in-memory or on-disk interactions), a user who can input messages to channels (by I/O stream), a network protocol that passes messages to channels (by port), etc. A skeleton program integrates built-in modules to support the implementation of external channels according to concrete requirements.

Notably, execution needs to take care of waiting for a channel. An internal channel $c$ gets into \textbf{waiting} if $c$ is synchronous and the sending system is not in the state right before sending a message or $c$ is asynchronous and $\textit{Cap}(c) < 1$. If $c$ is an external channel, it gets into waiting if no outside process sends a message to $c$.

\begin{example}[Channel in Transaction Client]
The transaction client in Example~\ref{eg:tx_client} has an anonymous action $c?\textit{tx}$ attached to the transition relation between \textit{init} and \textit{waiting}. This communication action will not proceed, i.e., channel $c$ gets into waiting, until consume a message via channel $c$. From the perspective of implementation, the execution only resumes when state variable $\textit{tx}$ has the value received from channel $c$.

In our implementation, information about a new transaction is pushed into channel $c$ when that transaction is submitted.
\end{example}

\subsection{Integration Stage}
\label{sec:integration_stage}
\textit{Integration Stage} serves the overall bottom-up approach that embeds or integrates the input system graph into a higher-level system graph, which is an incremental process. This stage also determines the next move to continue the iteration or produce a delivery.

\subsubsection{Increment}
An increment has two types: horizontal increment and vertical increment. A \textbf{horizontal increment} is to embed a system graph into another one. Formally, embedding is defined as follows.

\begin{definition}[Embedding]
\label{def:embedding}
Let $\mathfrak{S}_i = \langle D_i, \mathcal{N}_i, A_i, {\hookrightarrow}_i, i_i, g_{0,i}, F_i, P_i, \mathcal{L}_i \rangle, i \in [1,2]$ be two system graphs over $(\textit{Var}, \textit{Chan})$. System graph $\mathfrak{S}$ of embedding $\mathfrak{S}_1$ into $\mathfrak{S}_2$ in the place of state declarator $d_2 \in D_2$ is the tuple 
\[
    \langle D, \mathcal{N}, A, {\hookrightarrow}, i, g_0, F, P, \mathcal{L} \rangle
\]
where
\begin{itemize}
    \item $D = D_1 \uplus D_2$, $A = A_1 \uplus A_2$, $P = P_1 \uplus P_2$,
    \item $\forall d \in D_i: \mathcal{N}(d) = \mathcal{N}_i(d)$,
    \item ${\hookrightarrow} = {\hookrightarrow}_1 \uplus {\hookrightarrow}_2 \setminus {\hookrightarrow}_{d_2} \uplus {\hookrightarrow}_{d_2}'$,
    \item $i= \begin{cases} 
      i_1 & d_2 = i_2 \\
      i_2 & d_2 \neq i_2
   \end{cases}$,  $g_0= \begin{cases} 
      g_{0,1} & d_2 = i_2 \\
      g_{0,2} & d_2 \neq i_2
   \end{cases}$,
   \item $F= \begin{cases} 
      F_2 \setminus d_2 \uplus F_1 & d_2 \in F_2 \\
      F_2 & d_2 \notin F_2
   \end{cases}$,
   \item $\forall s \in S_i: \mathcal{L}(s) = \mathcal{L}_i(s)$.
\end{itemize}
${\hookrightarrow}_{d_2} \in {\hookrightarrow}_2$ is a set of transition relations such that
\begin{enumerate}
    \item $\forall (d_2, g, a, d_2') \in {\hookrightarrow}_2: d_2 \xhookrightarrow{g \downarrow a} d_2' \in {\hookrightarrow}_{d_2}$, and
    \item $\forall (d_2', g, a, d_2) \in {\hookrightarrow}_2: d_2' \xhookrightarrow{g \downarrow a} d_2 \in {\hookrightarrow}_{d_2}$.
\end{enumerate}

${\hookrightarrow}_{d_2}'$ is a set of transition relations such that
\begin{enumerate}
    \item $\forall (d, g, a, d_2) \in {\hookrightarrow}_2: d \xhookrightarrow{g_{0,1} \downarrow a} i_1 \in {\hookrightarrow}_{d_2}'$, and
    \item $\forall f \in F_1 (\forall (d_2, g, a, d) \in {\hookrightarrow}_2: f \xhookrightarrow{g \downarrow a} d)$.
\end{enumerate}
 
\end{definition}

\begin{remark}[Module]
\label{rem:module}
In Definition~\ref{def:embedding}, if $\mathfrak{S}_1$ shares state declarators, actions, propositions, naming and labeling functions with $\mathfrak{S}_2$, i.e., $D_1 \subseteq D_2, A_1 \subseteq A_2, P_1 \subseteq P_2, \forall d \in D_1: \mathcal{N}_1(d) = \mathcal{N}_2(d), \forall s \in S_1: \mathcal{L}_1(s) = \mathcal{L}_2(s)$, then $\mathfrak{S}_1$ is a module of $\mathfrak{S}_2$.

While embedding $\mathfrak{S}_1$ into $\mathfrak{S}_2$, if $\mathfrak{S}_1$ is a module of $\mathfrak{S}_2$, $\uplus$ relation is changed to $\cup$.
\end{remark}

A \textbf{vertical increment} is an integration of a system graph into another one through parallelisms or communications, i.e., the current system graph is regarded as a subsystem that is parallel with or communicates with other subsystems in a higher-level system.

\begin{example}[Increment in Transaction Client]
\label{eg:tx_client_increment}
For $\mathfrak{S}_{tx}$, \textit{pending} is a module. In fact, $\mathfrak{S}_{tx}$ is a considerably high-level model. Besides \textit{pending}, other components also encapsulate either a horizontal increment or a vertical increment. 
\end{example}

\subsubsection{Next Move}
As the final stage of an iteration, \textit{Integration Stage} determines the next move according to the current system graph $\mathfrak{S}$. If $\mathfrak{S}$ does not include all details in the design, then it is called \textbf{refinable}. Otherwise, it is \textbf{unrefinable}. If $\mathfrak{S}$ is not integrated into any other system graph, then it is called \textbf{independent}. Otherwise, it is \textbf{dependent}.

\begin{itemize}
    \item If $\mathfrak{S}$ is independent and unrefinable, then terminate its iterative and incremental process and deliver its implementation.
    \item If $\mathfrak{S}$ is dependent and unrefinable, then integrates $\mathfrak{S}$ into a higher-level system graph $\mathfrak{S}'$, and start the iterative process of $\mathfrak{S}'$.
    \item If $\mathfrak{S}$ is refinable, always go to the next iteration of $\mathfrak{S}$.
\end{itemize}

\section{Seniz}
\begin{figure}[htbp!]
\centering    
\includegraphics[width=9cm]{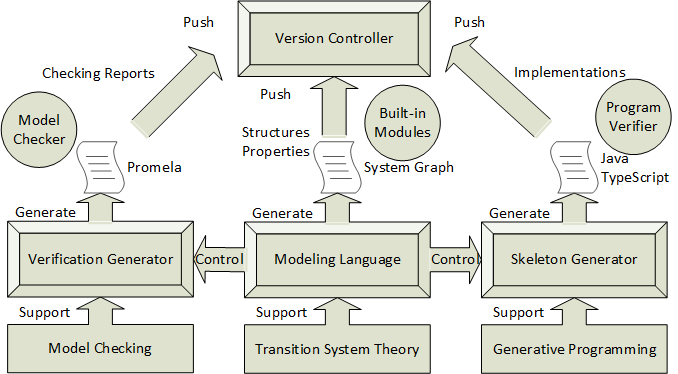}
\caption[Architecture]{Seniz architecture.}
\label{fig:architecture}
\end{figure}

We implement Seniz \footnote{https://github.com/yepengding/Seniz}, a framework that practicalizes FDD. As shown in Figure~\ref{fig:architecture}, Seniz consists of a modeling language, a verification generator, a skeleton generator, and a version controller.

The modeling language is used to abstract system graphs from real-world systems. It allows developers to formulate static structures by state declarators, dynamic changes by actions and transition relations, as well as expected system behaviors by formal propositions and properties.

The verification generator translates Seniz programs developed in \textit{Abstraction Stage} into Promela programs for \textit{Verification Stage}. Generally, a state declarator is translated into a \textit{macro} and an \textit{inline}. A \textit{macro} is defined as the conjunction of all state variables identified in the declarator and additional critical variables inferred by the compiler. The translation of propositions and temporal properties is trivial on account of the similar syntax.

Besides, the current version of Seniz supports generating Java and TypeScript programs from Seniz programs. The core method is illustrated in Section~\ref{sec:impl_stage} and follows the OOP paradigm. The generator integrates a powerful toolchain to support the functionality of the generated SDK and allows customization of the tech stack. The FDD developers only need to focus on the implementation of action effects.

Additionally, a rigorous version controller is mechanized in Seniz. Based on the rigorous definition of iterations and increments illustrated in Section~\ref{sec:integration_stage} and cryptographic hash function, the version controller automatically archives iterations and increments and labels them with verified properties.

\section{Case Study}
We conducted case studies on the development of consensus mechanisms, one of the core mechanisms in decentralized systems, to evaluate FDD in practice with the support of Seniz.

\subsection{Criteria}
\subsubsection{Design Quality}
Designs are correctly represented, as well as correctly and completely reflect requirement specifications and provably conform to implementations.

\subsubsection{Correctness}
Deliveries satisfy functionalities specified in requirements and function well.

\subsubsection{Productivity}
Developers keep a high efficiency during development and deliver verified products on time.

\subsection{Method}
We recruited 4 professional developers who have a similar level of experience in software design and programming but do not have experience in consensus mechanisms and formal methods. All participants are males and in the 25-28 age range. Besides, we collaborated with business project developers of Dagbase \cite{ding_dagbase_2020} to provide introductory lectures on consensus algorithms, the Dagbase project, FDD, and Seniz.

Our study was conducted in a remote setting. Based on the results of a self-efficacy questionnaire and programming skill test, we divided 4 participants into two groups $G_0$ and $G_1$ according to their preferred programming languages and tested skills. One of the authors also formed a group labeled as $G_2$. Each group is required to develop three simplified types of consensus mechanisms in order: Proof-of-Work (PoW), Hashgraph (HG), and Raft. The consensus mechanism in the \textit{Persistence Layer} of the Dagbase project is displaceable if the interfaces of the new mechanism satisfy the Dagbase specification. Hence, each group is also required to deliver an implementation that satisfies the defined interfaces. Besides, design documents are also required during the delivery.

$G_0$ has freedom of choice for their development, including process model, design method, programming paradigm, and tool, except FDD and Seniz. On the contrary, $G_1$ and $G_2$ are mandatorily required to use FDD as their process model and Seniz as their only tool.

\subsection{Measurements}
\subsubsection{Design Quality}
We use a qualitative method to measure design quality due to the difficulty of formulating a unified rubric for different design methods. Our qualitative method evaluates: 1) design is correct in its language, 2) design satisfies requirement specification, 3) implementation satisfies design, 4) design is well documented with clarity, completeness, understandability, and reproducibility.

\subsubsection{Correctness}
A quantitative method is adopted to score correctness based on how many test cases (including edge cases) the delivered implementation can pass.

\subsubsection{Productivity}
Participants select 4 time slots (3 hours per slot) for each consensus mechanism. After each time slot, a percentage number is reported to show the current progress.

\subsection{Results}
$G_0$ adopted a variant test-driven development process with UML as the design language, Java as the implementation language, object-oriented programming as the paradigm. $G_1$ and $G_2$ followed FDD with the Seniz support.

\subsubsection{Design Quality}
\begin{enumerate}
    \item[$G_0$] We found 11 mistakes in class diagrams, 8 mistakes and 3 unusual representations in sequence diagrams, 4 mistakes in state diagrams. The design only defines the main control flow in the requirement specification via sequence and state diagrams, while most flow paths are left undefined and unverified. Furthermore, the design fails to define formal properties implied from the requirement specification. Implementations satisfy the design, but the design lacks some critical details such as sub-process and concurrent interactions.
    \item[$G_1$] Design is correct, ensured by the Seniz language compiler. However, we found 10 warnings about the unverifiable properties, which implies 10 formal properties may not hold and satisfy the requirement specification. Except that, 39 formal properties are verified. The design fully defines the control flow. Implementations are proved to satisfy the design with critical details.
    \item[$G_2$] Almost the same evaluation with $G_1$ except we have 8 warnings about the unverifiable properties and 61 formal properties are verified.
\end{enumerate}

\subsubsection{Correctness and Productivity}
The evaluation result is shown in Table~\ref{tb:eval_res}. Column \textit{C (*)} is the correctness evaluation with the total number of test cases in the parenthesis, while \textit{P} denotes the productivity evaluation scaled from 0 to 10. If the final number in \textit{P} is smaller than 10, it means the group failed to finish the mechanism after consuming all time slots.

\begin{table}[htbp]
\caption{Evaluation Results of Correctness and Productivity}
\label{tb:eval_res}
\begin{center}
\begin{tabular}{|c|c|c|c|c|c|c|}
\hline
\textbf{Group}&\multicolumn{2}{|c|}{\textbf{PoW}} & \multicolumn{2}{|c|}{\textbf{Hashgraph}} & \multicolumn{2}{|c|}{\textbf{Raft}} \\
\cline{2-7}
 & \textit{C (33)} & \textit{P} & \textit{C (56)} & \textit{P} & \textit{C(49)} & \textit{P} \\
\hline
$G_0$ & 33 & 3/5/8/10 & 44 & 1/4/7/10 & 49 & 4/7/9/10 \\
$G_1$ & 31 & 1/2/5/7 & 53 & 1/3/5/7 & 49 & 1/3/6/10 \\
$G_2$ & 33 & 1/3/7/10 & 56 & 1/3/7/9 & 49 & 2/4/8/10 \\
\hline
\end{tabular}
\end{center}
\end{table}

\subsection{Analysis}
From the \textit{Design Quality} result, we find that it is hard to use UML diagrams to define the full control flow in a limited time, and manual design gets mistakes easily. Besides, due to the lack of details to describe all kinds of aspects in design, many methods are "blindly" implemented without considering potential risks. For instance, an undesigned and misimplemented directed acyclic graph checking algorithm of $G_0$ caused forking in some edge cases, which is the main reason that caused the significant correctness difference in the \textit{Correctness} result of Hashgraph. However, with the specified formal property $\forall \Diamond (\textit{ForkFree} \land \textit{Ancestor}(e_1,e_0) \to \forall \Box \textit{See}(e_0,e_1))$ of $G_2$, the forking issue can be located before the implementation.

The failed 2 test cases of $G_0$ in the PoW case are caused by unfamiliarity with the Seniz built-in hashing library. The failed test cases in the Hashgraph implementation are caused by the improper abstraction of the \textit{Coin process}, which made the consensus fail to terminate. We also find that unverifiable properties are associated with either probabilistic properties or divergence branches.

The productivity of $G_0$ is significantly higher than $G_1$ and $G_2$ at the start, implying that it requires more time to find a suitable start point and produce a preliminary milestone. Based on the survey results, it is hard to formulate the first abstraction for a system in FDD, and improper abstractions will lead to difficulties for later refinements. Besides, both $G_1$ and $G_2$ have unfinished circumstances. However, it is not technically fair without considering the productivity of rigorous designs.

Notably, these case studies are limited to the scope of consensus mechanisms and cannot fully demonstrate the effectiveness of FDD in other components and mechanisms of decentralized systems.

\section{Related Work}
\subsection{Verification-Driven Engineering}
Verification-driven engineering (VDE) \cite{kordon_model_2008} integrates the formal methods in MDD and particularly promotes formal verification during the development process. In \cite{kordon_model_2008}, the principles and requirements for better use of formal methods in MDD are illustrated. It concludes that it is necessary to switch from MDD to VDE even though it still needs more sophisticated techniques to support it.

Some works also enhance the MDD by introducing verification-driven methods such as \cite{shaikh_verification-driven_2010}. It proposes a verification-Driven slicing technique to partition the model into submodels while preserving properties by formal verification. Sphinx \cite{mitsch_collaborative_2014} is proposed as a VDE toolset for modeling and verifying hybrid systems. It defines semantics for the UML activity diagrams. In \cite{menghi_verification-driven_2019}, a verification-driven framework named FIDDle is presented and evaluated by developing parts of the K9 Mars Rover model.

However, none of them tackle the problem from the perspective of an agile development process. The relationships between models also lack formalization. FDD formally defines different stages and introduces a formalism to manage the process rigorously.

\subsection{Blockchain-Oriented Engineering}
Recently, the research on the blockchain-oriented development process has made some progress.

The work \cite{dittmann_model-driven_2019} studies three approaches to model and implement a taxi dispatcher application on a blockchain, including an extended BPMN approach, using synchronized state-machines, and high-level Petri nets. In \cite{xu_model-driven_2019}, a code generation method for smart contracts is proposed based on MDD for collaborative business processes.

Besides MDD, other methods are also studied to optimize the development process of blockchain applications such as \cite{marchesi_agile_2018,wessling_engineering_2018}. In \cite{marchesi_agile_2018}, it proposes an agile software engineering method to organize the development process with concrete plans and introduces a set of new UML stereotypes to enhance the modeling capability. Some architectural patterns extracted from existing decentralized applications are studied in \cite{wessling_engineering_2018}.

Notably, they focus on providing application-level solutions for the development of decentralized applications and barely introduce formal methods as a critical component in their methodologies.

\section{Conclusion}
In this paper, we proposed FDD, a novel iterative and incremental development process for developing provably correct decentralized systems with formal methods. Besides, we presented a framework named Seniz to practicalize FDD. Furthermore, we demonstrated the effectiveness of FDD and Seniz in practice by conducting case studies. In the meantime, Seniz still has much room for improvement, including efficiency, generated code quality, and user interface. Additionally, the scope of FDD can be extended to post-quantum DLTs and other types of security-sensitive systems.

%
%
%
\bibliographystyle{splncs04}
\bibliography{mybibliography}

\end{document}